\documentclass[3p,times,twocolumn]{elsarticle}

\pdfoutput=1 
\usepackage{natbib,ecrc}
\newcommand{\apjl}{ApJ Letter}
\newcommand{\apj}{ApJ}
\newcommand{\aap}{A\&A}
\newcommand{\mnras}{MNRAS}
\newcommand{\lsi}{LS~I~+61$^{\circ}$303}
\newcommand{\ls}{LS\,5039}
\newcommand{\hess}{HESS J0632+057}
\newcommand{\fermi}{{\it Fermi}}

\volume{00}

\firstpage{1}

\journalname{Nuclear Physics B Proceedings Supplement}

\runauth{G. A. Caliandro and A. B. Hill}

\jid{nuphbp}

\jnltitlelogo{Nuclear Physics B Proceedings Supplement}

\usepackage{amssymb}
\usepackage[figuresright]{rotating}

\begin{document}

\begin{frontmatter}



\dochead{}

\title{The strange case of HESS J0632+057 and the $\gamma$-ray High Mass X-ray Binaries}


\author{G. A. Caliandro[1] and A. B. Hill[2][3], on behalf of the \fermi-LAT collaboration}

\address[1]{Institut de Ciencies de l'Espai (IEEC-CSIC), Campus UAB, 08193 Barcelona, Spain}
\address[2]{W. W. Hansen Experimental Physics Laboratory, Kavli Institute for Particle Astrophysics and Cosmology, Department
of Physics and SLAC National Accelerator Laboratory, Stanford University, Stanford, CA 94305, USA}
\address[3]{School of Physics and Astronomy, University of Southampton, Highfield, Southampton, SO17 1BJ, UK}

\begin{abstract}
In the last decade Cherenkov telescopes on the ground and space-based $\gamma$-ray instruments have identified a new class of high mass X-ray binaries (HMXB), whose emission is dominated by $\gamma$ rays. To date only five of these systems are known. All of them are detected by Cherenkov telescopes in the TeV energy range, while at GeV energies there is still one (HESS J0632+057) that has no reported detection with the \fermi-LAT. A deep search for $\gamma$-ray emission of HESS J0632+057 has been performed using more than 3.5 years of \fermi-LAT data. We discuss the results of this search and compare it to other $\gamma$-ray binary systems.
\end{abstract}


\end{frontmatter}

% \linenumbers
\section{Introduction}
\label{intro}

The High Mass X-ray Binaries (HMXBs) are
relatively young ($<10^8$ year) Galactic sources composed of a massive OB or Be type star and a compact object,
either a neutron star or a black hole. These systems are  bright X-rays emitters, 
generally (but not always) due to matter from the massive star accreting onto the compact object.

In the last decade a handful of HMXBs have been detected at high (HE;
0.1--100\, GeV) or very high-energies (VHE; $>$100 GeV). They are 
LS~I~+61$^{\circ}$303 \citep{2006Sci...312.1771A, 2008ApJ...679.1427A, 2009ApJ...701L.123A}, LS\,5039 \citep{2005Sci...309..746A,2009ApJ...706L..56A}, PSR B1259$-$63 \citep{2005A&A...442....1A,B1259,B1259Tam},
1FGL J1018.6$-$5856 \citep{1FGLbinary_science}, Cyg X$-$3 \citep{2009Sci...326.1512F},  and Cyg X$-$1 \citep{2007ApJ...665L..51A,2010ApJ...712L..10S}. With the exception of the last two sources, for the others the power of their emission is dominant at high energies, rather than at longer wave-lengths. For this reason they are some times mentioned with the name of $\gamma$-ray HMXBs.
Recently a new source has been claimed to belong to this class of sources. It is HESS J0632+057.
In this proceeding we summarize the path covered from the discovery of \hess\ as unidentified source, to the claiming of its binarity.
Finally, we describe the deep search for its emission in the GeV energy range, and we discuss the analogies and the differences that this particular source shows with respects the others $\gamma$-ray HMXBs.

\begin{figure*}[th]
\center
\includegraphics[width=0.45\textwidth]{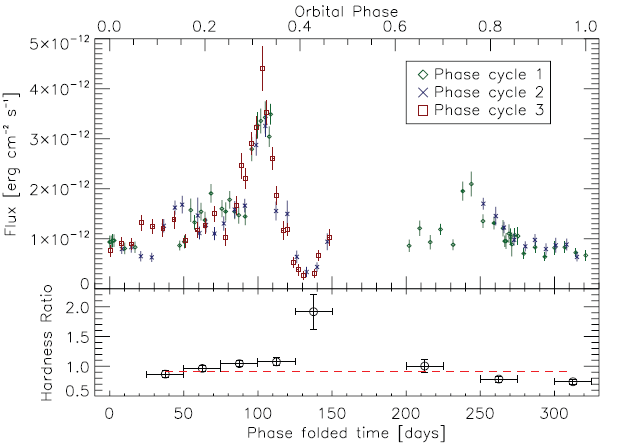}
\hspace{0.5 cm}%
\includegraphics[width=0.45\textwidth]{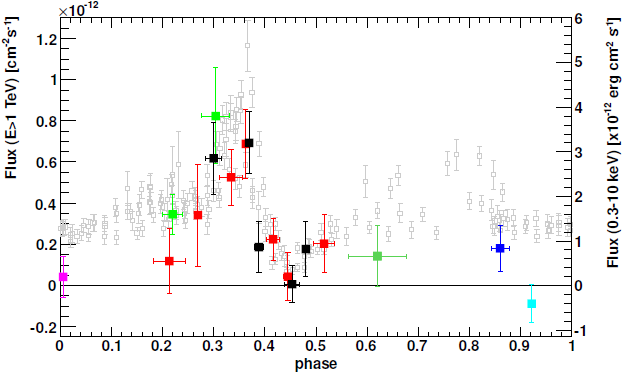}
\caption{{\em Left}: Figure from \citep{Bongiorno2011}. X-ray light curve of \hess\ folded over the period of 321 days. The lower panel shows the hardness ratio (2.0–10.0 keV)/(0.3–2.0 keV), folded over the same period, and binned at 25 day intervals.
{\em Right}: Figure from \citep{2012arXiv1210.4703M}. Integral $\gamma$-ray fluxes above 1 TeV of HESS J0632+057 from VERITAS (filled
markers) and Swift XRT X-ray measurements (0.3-10 keV) in gray. The colors indicate the different observing periods of VERITAS.
\label{foldedLC}}
\end{figure*}

\section{Discovery, and Cherenkov monitoring}
\label{discovery}
HESS J0632+057 was discovered at VHE energies by the High Energy Spectroscopic System (H.E.S.S.) on 2007 \citep{hess_discovery}. It is in the Monoceros Loop region, very close to the Rosette Nebulae. It was classified as a point-like unidentified source. It is interesting to note that all the VHE Galactic sources are extended, with the exception of the HMXBs. Therefore, the point-like feature of HESS J0632+057 was a strong indication of its binary nature. Further more, of many tens of unidentified sources discovered in the Galactic Plane H.E.S.S. survey, HESS J0632+057 is one of the only two unidentified VHE sources which appears to be point-like within the experimental resolution of the instrument. The other (HESS J1745-290) is coincident with the gravitational center of the Milky Way \citep{2006PhRvL..97v1102A}. 
When HESS J0632+057 was discovered, several plausible association with sources in other wave-lengths were discussed. A posteriori, the most relevant was the association with a massive emission-line star of spectral type B0pe (MWC 148, see section \ref{OptRadio}).
The flux and the spectral index of this source measured by H.E.S.S. are $F(E>1 {\rm TeV}) = 6.4 \pm 1.5 \times 10^{-13}$ cm$^{-2}$ s$^{-1}$, $\Gamma = 2.53 \pm 0.26$, respectively.

Subsiquently to the discovery, VERITAS and MAGIC started to monitor HESS J0632+057. It was observed by VERITAS for 31 hr in 2006, 2008, and 2009. The first observation was pointed toward the center of the Monoceros region, (at an angular distance of $\sim 0^{\circ}.5$ from the HESS J0632+057). The other observations were targeted around the reported position of HESS J0632+057. During these observations, no significant signal in at energies above 1 TeV was detected, and a flux upper limit of $4.2 \times 10^{-13}$ cm$^{-2}$ s$^{-1}$ at 99\% confidence level \citep{veritas_ul}.  
The non detection by VERITAS during these observations provides variability in the VHE flux of HESS J0632+057. This was in favor with the binary interpretation of this source.

Afterwards, the source was detected by MAGIC and VERITAS in February 2011, when TeV observation were triggered by an outburst in X-rays \citep{falcone2011,veritas_icrc_2011,magic2012}. The spectrum measured by MAGIC is compatible with the one measured by H.E.S.S. Further observations with VERITAS have been performed until February 2012 \citep{2012arXiv1210.4703M}. In the right panel of figure \ref{foldedLC} are plotted the integral $\gamma$-ray fluxes above 1 TeV of HESS J0632+057 measured by VERITAS in the different observations.

\section{X-ray follow up, and orbital period detection}
\label{Xray}

Since its discovery, HESS J0632+057 was followed by several X-ray observatories. As first, an observation taken
on the 2007 September 17 with the EPIC camera of {\em XMM-Newton} lead to the detection of a new X-ray source (XMMU J063259.3+054801) within the error box of HESS J0632+057, and coincident with the massive star MWC 148. This source exhibits a hard spectrum, consistent with an absorbed power law with $\Gamma = 1.26 \pm 0.04$, and the unabsorbed $1-10$ keV flux is $(5.3 \pm 0.4) \times 10^{-13}$ erg cm$^{-2}$ s$^{-1}$. 

A long monitoring of this source was planned with the {\em Swift} X-ray telescope. A first set of data covering 108 days was analyzed by \citep{falcone2010}. They measured flux increases by a factor $\sim 3.$, and found that X-ray variability was present on multiple timescales including days to months.  

The X-ray outburst on February 2011 that was announced by \citep{falcone2011} triggered the observation of VERITAS and MAGIC (see section \ref{discovery}), as well as an observation with {\em Chandra} requested by \citep{reatorres2011}. Comparing the results of the Chandra data analysis with those of the first observation of {\em XMM-Newton} \citep{reatorres2011} found also a a spectral variability in addition to the flux one. Indeed, the spectral index measured was $1.61 \pm 0.03$. 
The flux and spectral variability are both characteristic features of the HMXBs, but in contrast with what observed for the other TeV binaries, in this source the higher the flux the softer the X-ray spectrum. 

Finally, the detection of the orbital periodicity of $321 \pm 5$ days was claimed by \citep{Bongiorno2011}. The periodical X-ray flux modulation was the result of the analysis of the long set of {\em Swift} observations of the source from 2009 January to 2011 March. In figure \ref{foldedLC} (left plot) is shown the folded X-ray light curve of HESS J0632+057. It shows a narrow high peak at phase 0.3 followed by a dip feature, and centered at phase 0.75 there is a moderate broad peak. 
The {\em XMM-Newton} observation fall in the dip feature, where the spectrum is harder, while the {\rm Chandra} observation was taken during the high peak. Phase 0.0 in this plot is arbitrary set at the beginning of the {\em Swift} observations. 

\section{Optical and Radio counterparts}
\label{OptRadio}

MWC 148 is the companion massive star in the binary system HESS J0632+056. It is a  Be star, as in the case of the HMXB systems \lsi, and PSR B1259-63. Before the detection of the orbital period in X-rays, \citep{aragona2010} analyzed optical spectra of the star acquired  to investigate the stellar parameters, the properties of the Be star disk, and evidence of the binarity. They derived a mass of 13.2-19.0 $M_{\odot}$, and a radius of 6.0-9.0 $R_{\odot}$. Fitting the spectral energy distribution they found a distance between 1.1 and 1.7 kpc. Since their dataset was too short ($\sim 1$ month) it was not possible to detect the orbital period of the system. On the other hand, \citep{casares2012} collected optical spectra observations spanning for 4 years from 2008 October to 2011 May. They calculated the orbital parameters of the binary systems, finding that it is highly eccentric with $e = 0.83 \pm 0.08$. Furthermore, the orbital solution implies that the high peak in the X-ray light curve is delayed $\sim 0.3$ orbital phases after the periastron passage, similar to the case of \lsi.

In the radio regime \cite{skilton2009} detected with the Giant Metrewave Radio Telescope (GMRT) and the Very Large Array (VLA)
observations at 1.28 and 5\,GHz, respectively, a point-like, variable radio source at the position of MCW 148. \citep{moldon2011} observed \hess\ with the European VLBI Network (EVN) at 1.6\,GHz in two epochs: during the January/February 2011 X-ray outburst and 30 days later. In the first epoch the source appears point-like, whereas in the second one it appears extended with a projected size of $\sim$75\,AU assuming a distance to the system of 1.5\,kpc. The brightness temperature of 2$\times$10$^{6}$\,K at 1.6\,GHz together with the negative spectral index around \mbox{$-$0.6} point to non-thermal synchrotron radiation producing the radio emission \citep{skilton2009,moldon2011}. The detected projected displacement of the peak of the emission of 21\,AU in 30 days, the morphology and the size of \hess\ are similar to the ones observed in the well-established $\gamma$-ray binaries PSR B1259$-$63, \ls, \lsi.

\section{Search for GeV emission}
\label{GeV}
All the $\gamma$-ray HMXBs have a bright emission in the GeV energy range, but \hess\ is not present in the Second \fermi-LAT Source Catalog \citep{2FGL} (2FGL hereafter). Therefore, we performed a fine analysis with the \fermi-LAT data in order to deeply search for GeV emission.
A dataset of $\sim 3.5$ years from $4^{th}$ August 2008 to $3^{rd}$ March 2012 has been analyzed, that is almost twice the dataset used for the 2FGL.

\begin{figure}
\center
\includegraphics[width=0.45\textwidth]{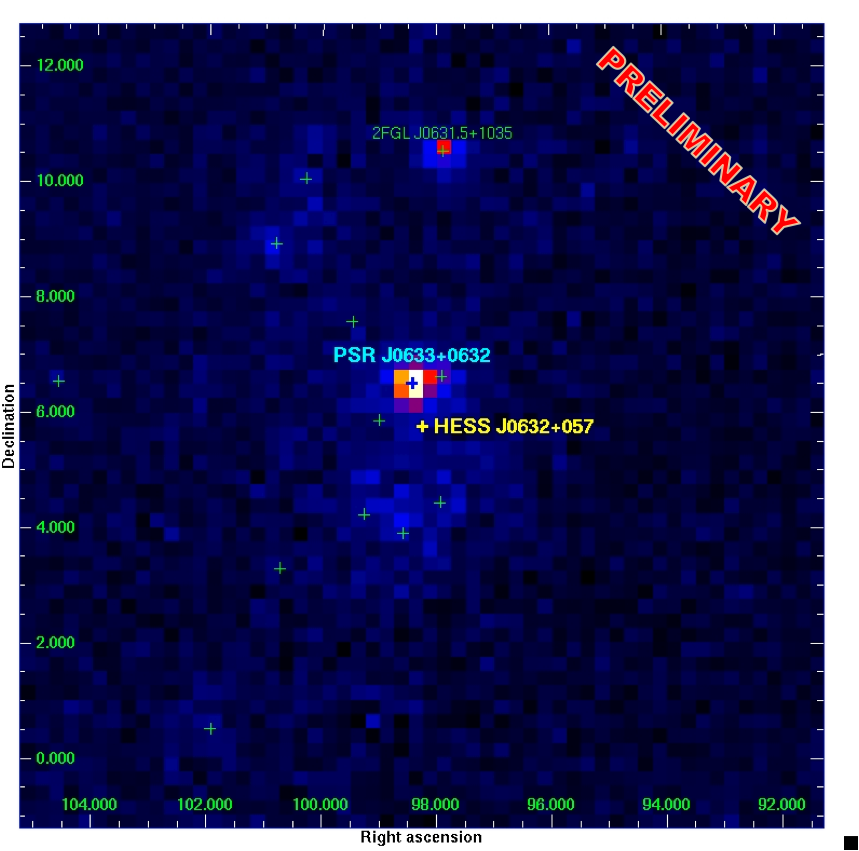}
\caption{Background subtracted count map above 1 GeV of the $10^{\circ} \times 10^{\circ}$ sky region centered on HESS J0632+057. The pixel size is $0.1^{\circ} \times 0.1^{\circ}$. A Gaussian smoothing with 3 pixel kernel radius is applied. The 2FGL sources are labeled with a cross, HESS J0632+057 by the diamond, and PSR J0633+0632 by the circle.
\label{Cmap}}
\end{figure}

HESS J0632+057 lies at $b=-1.44^{\circ}$ from the Galactic plane, in the active region of the Monoceros loop.
The region at GeV energies appears as shown in Figure \ref{Cmap}, 
where the sources from the 2FGL catalog are labeled with a green cross, the yellow one in the center marks the position of HESS J0632+057, while in cyan is marked the pulsar PSR J0633+0632. 
It is evident that the pulsar emission dominates the region aroung \hess. Their angular distance is only 0.78$^{\circ}$. 

\subsection{PSR J0633+0632}

\begin{figure}
\center
\includegraphics[width=0.45\textwidth]{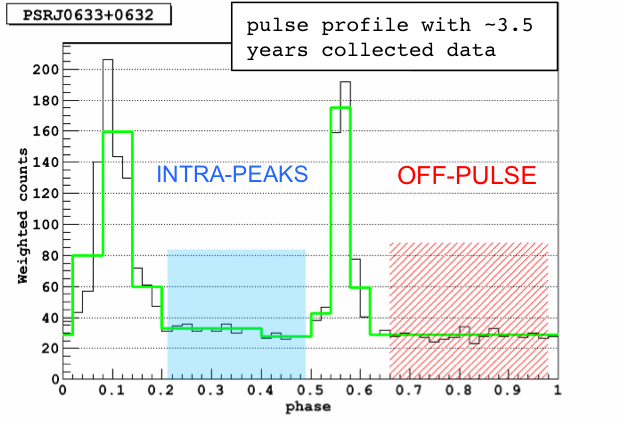}
\caption{Weighted pulse profile of PSR J0633+0632 with $\sim 3.5$ years collected \fermi-LAT data. The weight associated to each event count corresponds to the probability that the event is emitted by the pulsar, rather than by a nearby source or by the diffuse emission. The off-pulse, and the intra-peaks phases are defined by mean of a Bayesian block algorithm (green line): (0.66 - 0.98) and (0.21 - 0.49), respectively.
\label{PSRJ0633}}
\end{figure}

PSR J0633+0632 is a bright radio-quiet $\gamma$-ray pulsar discovered in the first six months of the \fermi\, mission \citep{16blind}. Indeed, it is listed in the first \fermi-LAT catalog of $\gamma$-ray pulsars \citep{1PC}.
It has a flux of $8.4 \pm 1.2 \times 10^{-8}$ ph cm$^{-2}$ s$^{-1}$, and the pulse profile has the characteristic two narrow peaks separated $\sim 0.5$ in phase. The weighted pulse profile of PSR J0633+0632 is shown in figure \ref{PSRJ0633}. 
A weight was associated to each event, corresponding to the probability that the event is emitted by the pulsar, rather than by a nearby source or by the diffuse emission. 
We defined the off-pulse phases of PSR J0633+0632 (0.66 - 0.98) applying a Bayesian block algorithm adapted for weighted-counts light curves \citep{CaliandroHill}. In figure \ref{PSRJ0633} the green line shows the calculated Bayesian blocks. The off-pulse is defined as the lowest block reduced by 10\% of its total length at each edge. As well, we defined the intra-peaks or bridge phases (0.21 - 0.49).

In order to avoid the contamination from the strong emission of PSR J0633+0632, we carried on the analysis of HESS J0632+057 selecting the off-pulse and the intra-peaks phases of PSR J0633+0632.

With this phase selection the regioin around \hess\ appears quite crowded, and not perfectly modeled by the 2FGL sources. 
This is mainly due to the fact that the 2FGL catalog is built with 2 years of \fermi-LAT data, while in this analysis the dataset is almost doubled ($\sim 3.5$ years).

\begin{figure}
\center
\includegraphics[width=0.45\textwidth]{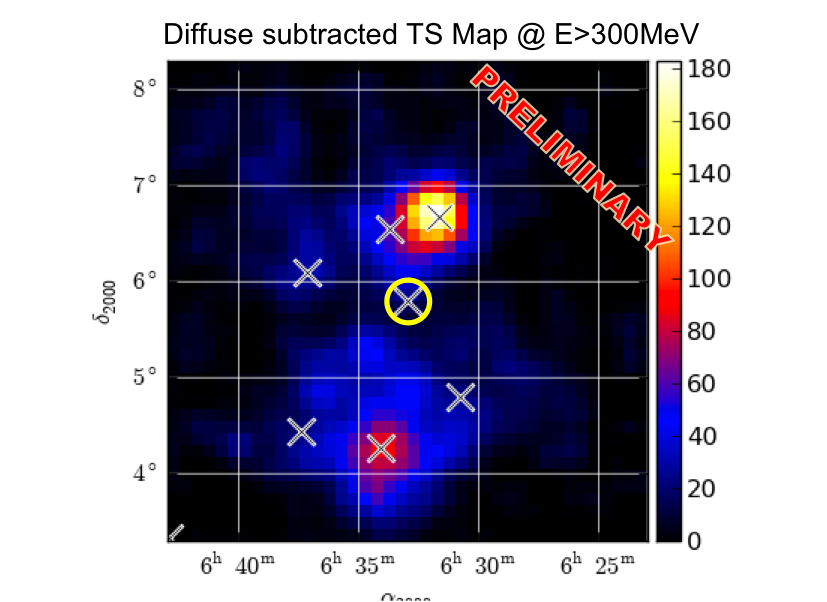}
\caption{Diffuse subtracted TS map of the $5^{\circ} \times 5^{\circ}$ sky region centered on HESS J0632+057. The maps are calculated for $E > 300$ MeV. The 2FGL sources are labeled with a cross, HESS J0632+057 by the yellow circle.
\label{TSmaps}}
\end{figure}

Figure \ref{TSmaps} shows the region out of the pulsar peaks. The 2FGL sources are labeled with a cross, and HESS J0632+057 by the yellow circle.
Before to analyze directly \hess\, we as first applied an iterative method to reach a good modeling of the region around it. The method developed is described in detail in \citep{CaliandroHill}. It accounts for the excesses in the map adding new point-like sources in the model of the region.

\subsection{Upper limits of \hess}
Once we obtained a good model for the region, we came back to focus specifically on the analysis of HESS J0632+057.
We found no significant detection of the GeV emission of \hess. 
Then, we derived the 95\% flux upper limit for this source.  
The Helene's Bayesian method \citep{Helene} was adopted to evaluate them. We calculated the flux upper limit for energies $E>100$ MeV of $F_{100} < 3.0 \times 10^{-8}$ cm$^{-2}$ s$^{-1}$.

\subsection{Folded analysis and variability}

\begin{figure}
\center
\includegraphics[width=0.45\textwidth]{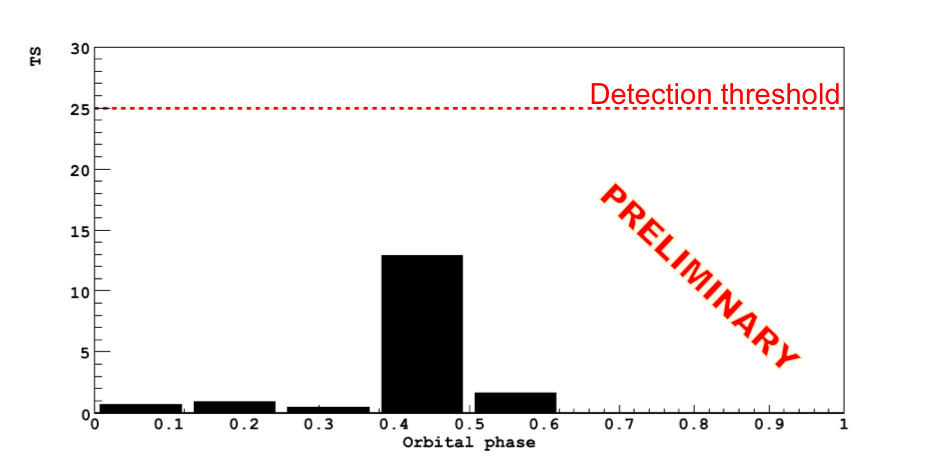}
\caption{The TS light curve of \hess\ calculated folding the \fermi-LAT data with the orbital period of 321 days. The TS values of all the phase bins are below the detection threshold (TS$=25$). 
\label{foldedTS}}
\end{figure}

In the analysis described so far we have considered HESS J0632+057 as a steady source, and we concluded that its average $\gamma$-ray emission is below the sensitivity of \fermi-LAT.  But it is still possible that the system have a brighter emission during a specific part of the orbit, like for the case of the HMXB system PSR B1259-63 (\citep{B1259}, \citep{B1259Tam}). 
In order to investigate this hypothesis, a further analysis was performed folding the \fermi-LAT data with the orbital period of 321 days, and setting the phase 0.0 to the epoch 54857 MJD \citep{Bongiorno2011}. We subdivided the orbit in 8 equally spaced phase bins, and for each of them we calculated the significance of the GeV emission. 
The results are showed in figure \ref{foldedTS}. We found that none of the bins have a significacnce above the detection threshold.
Finally, we also searched for flares or high flux states of HESS J0632+057, calculating light curves with different time binning scales, from days to months. For none of the time scales investigated we got a significant signal.

\section{Conclusions}
\label{Conclusions}

In this proceeding we have summarized all the steps leading to the claim that \hess\ is a $\gamma$-ray HMXB system.
\fermi-LAT data has been analyzed with the aim to search for the emission in the GeV energy range of \hess.
The analysis performed needed particular attention especially for the presence of a very nearby $\gamma$-ray pulsar 
PSR J0633+0632 (section 5.1). As first, we selected the events out of the peaks of the pulsars. 
Then, the region around the source was modeled adding new point-like sources in correspondence of unmodeled excesses with an iterative method.
Despite the good modeling of the region, we did not detected any significant signal of $\gamma$-ray emission from \hess.
We calculated the upper limit for its permanent GeV emission, $F_{100} < 3.0 \times 10^{-8}$ cm$^{-2}$ s$^{-1}$.
We also searched for GeV emission from a restricted portion of the orbit phases, 
as well as for plausible flares, or changes of state in its flux.

The lack of significant signal from all these searches is difficult to understand if compared with the behavior of the other 
$\gamma$-ray HMXBs in the same energy range. Similar to \hess, the massive stars in \lsi\ and PSR B1259-63 systems are Be star, that are characterized by the presence of a dense circumstellar disk.
The distances of these three HMXBs are comparable: $\sim 1.5$ kpc for \hess, $\sim 1.9$ kpc for \lsi, and $\sim 2.3$ kpc for 
PSR B1259-63. In contrast their orbital periods are significantly different.
\lsi\ has the shortest orbital period of about one month (26.5 days). As consequence, the systems is also the most compact among the three. In the GeV energy range, \lsi\ has a continuous emission modulated by its orbital motion. 
Most probably the GeV emission is present in every phase of the orbit because it is so small that the compact object is always interacting with the circumstellar disk, or with the stellar wind.
A different case is that of PSR B1259-63 with its very large orbital period of about 3.5 years. The orbit of this system is highly eccentric, with eccentricity $e \sim 0.87$. This makes that when the compact object is close to the periastron, it pass through the circumstellar disk. Only during this passage PSR B1259-63 emits GeV and TeV $\gamma$-rays.
\hess\ has an orbital period of about 1 year, that is in between those of \lsi\ and PSR B1259+63. 
The eccentricity of the orbit of \hess\ is also very high, with $e \sim 0.83$. We expected a GeV emission at least during the periastron passage, but we did not detected it. This can be attributed to different factors. An important role can be played by the features of the circumstellar disk, like its size, its density, and its inclination respect to both the observer line of sight and the plane of the orbit. On the other hand, the differences among \lsi\, PSR B1259+63, and \hess, can also be attributed to the features of the compact objects, like the intensity of their relativistic winds. Or they could be caused by the different nature of the compact object itself, either a pulsars (as for PSR B1259+63) or a black hole.
All these hypothesis are currently under investigation.\\
\\

{\textit{Acknowledgments:}}
The \fermi-LAT Collaboration acknowledges support from a number of agencies and institutes for both development and the operation of the LAT as well as scientific data analysis. These include NASA and DOE in the United States, CEA/Irfu and IN2P3/CNRS in France, ASI and INFN in Italy, MEXT, KEK, and JAXA in Japan, and the K.~A.~Wallenberg Foundation, the Swedish Research Council and the National Space Board in Sweden. Additional support from INAF in Italy and CNES in France for science analysis during the operations phase is also gratefully acknowledged. We thank the Spanish MICINN for additional support.

\end{document}